\documentclass[sigconf]{acmart}
\AtBeginDocument{%
  }

\setcopyright{acmlicensed}
\copyrightyear{2026}
\acmYear{2026}
\acmConference[KDD '27]{The 33rd ACM SIGKDD Conference on Knowledge
  Discovery and Data Mining}{August 09--13,
  2027}{San Jose, CA, USA}

\microtypesetup{expansion=true}
\usepackage{tikz}
\usepackage{pgfplots}
\pgfplotsset{compat=1.18}
\usepackage{enumitem}
\newcommand{\revfix}[1]{#1}
\usepackage{stfloats}
\usepackage{afterpage}
\usepackage{float}
\usepackage[most]{tcolorbox}
\usepackage{algorithm}
\usepackage{algorithmic}

\usetikzlibrary{positioning, arrows.meta, calc}
\newcommand{\zh}[1]{}
\begin{document}

\title{From Relevance to Utility: Rubric Alignment in Yuanbao Search}


\author{Yu Lu}
\affiliation[obeypunctuation=true]{%
  \institution{Tencent Inc., }
  \city{Beijing, }
  \country{China}
}
\email{luyuresearch@foxmail.com}

\author{Lixin Zou}
\correspondingauthor
\affiliation{%
  \institution{School of Cyber Science and Engineering, Wuhan University}
  \city{Wuhan}
  \country{China}
}
\email{zoulixin@whu.edu.cn}

\author{Yangqi Chen}
\affiliation[obeypunctuation=true]{%
  \institution{Tencent Inc., }
  \city{Beijing, }
  \country{China}
}
\email{yangqichen@tencent.com}

\author{Hui Xu}
\affiliation[obeypunctuation=true]{%
  \institution{Tencent Inc., }
  \city{Beijing, }
  \country{China}
}
\email{lucashxu@tencent.com}

\author{Qiaolin Xia}
\affiliation[obeypunctuation=true]{%
  \institution{Tencent Inc., }
  \city{Beijing, }
  \country{China}
}
\email{jolinxia@tencent.com}

\author{Hao Wang}
\affiliation[obeypunctuation=true]{%
  \institution{Tencent Inc., }
  \city{Beijing, }
  \country{China}
}
\email{vichwang@tencent.com}

\author{Qianwen Ma}
\affiliation[obeypunctuation=true]{%
  \institution{Tencent Inc., }
  \city{Beijing, }
  \country{China}
}
\email{qwenma@tencent.com}

\author{Reed Li}
\affiliation[obeypunctuation=true]{%
  \institution{Tencent Inc., }
  \city{Beijing, }
  \country{China}
}
\email{reedsli@tencent.com}

\author{Shuang Zeng}
\affiliation[obeypunctuation=true]{%
  \institution{Tencent Inc., }
  \city{Beijing, }
  \country{China}
}
\email{shuangzeng@tencent.com}

\author{Haijin Liang}
\affiliation[obeypunctuation=true]{%
  \institution{Tencent Inc., }
  \city{Beijing, }
  \country{China}
}
\email{hodgeliang@tencent.com}

\author{Haibo Shi}
\affiliation[obeypunctuation=true]{%
  \institution{Tencent Inc., }
  \city{Beijing, }
  \country{China}
}
\email{haiboshi@outlook.com}

\renewcommand{\shortauthors}{Y. Lu et al.}

\begin{abstract}
\looseness=-1
Agentic search demands that a search engine move beyond per-document
relevance ranking to assemble document sets optimized for downstream
LLM reasoning. 
Prior work addresses this either by improving
individual pipeline stages in isolation or by applying RL-based
selection on top of relevance-trained upstream modules, leaving the
full stack unaligned to utility for reasoning. 
We propose
rubric-aligned search, a framework that unifies the traditional search pipeline
under a single supervision surface. For each query, we derive a rubric from coarse to fine, specifying
weighted positive criteria that the document set must satisfy and negative
criteria that it must avoid. From this rubric we derive two training
signals. First, a lightweight labeler distills rubric-grounded
judgments into per-document utility labels, which supervise a
contrastive dual-encoder for retrieval and a multi-task ranker trained with four rubric-aligned scoring heads. Second, a rubric-conditioned evaluator
produces a set-level reward that drives a dual-branch RL selector,
combining GRPO with best-rollout distillation
to resolve credit assignment in subset optimization. 
Deployed in Yuanbao Search, the
pipeline improves end-to-end answer quality by 8.52\% over the online production baseline.
\end{abstract}


\settopmatter{authorsperrow=3}
\maketitle

\section{Introduction}


\looseness=-1 For decades, web search has been dominated by learning to
rank: given a query, order documents by relevance, authority, freshness,
and diversity~\cite{burges2010ranknet,yin2016ranking}. The goal is to maximize
per-position utility for a human user who scans the ranked list and
synthesizes information.
Recently, agentic search engines have changed this consumption
model~\revfix{\cite{lewis2020rag,curiosity2026}}: instead of browsing a ranked list,
the user receives a single answer synthesized by an LLM that reasons
over a bounded set of documents, synthesizing a multi-source summary or answering a multi-hop question that no single document covers, so the quality of its answer is entirely gated by the \emph{utility} of the document set it receives. The retrieval objective therefore shifts
	from ranked-list quality to \emph{set-level utility}: a compact, conflict-free document set that
	maximizes answer accuracy.

\begin{figure}[t]
\centering
\includegraphics[width=\columnwidth]{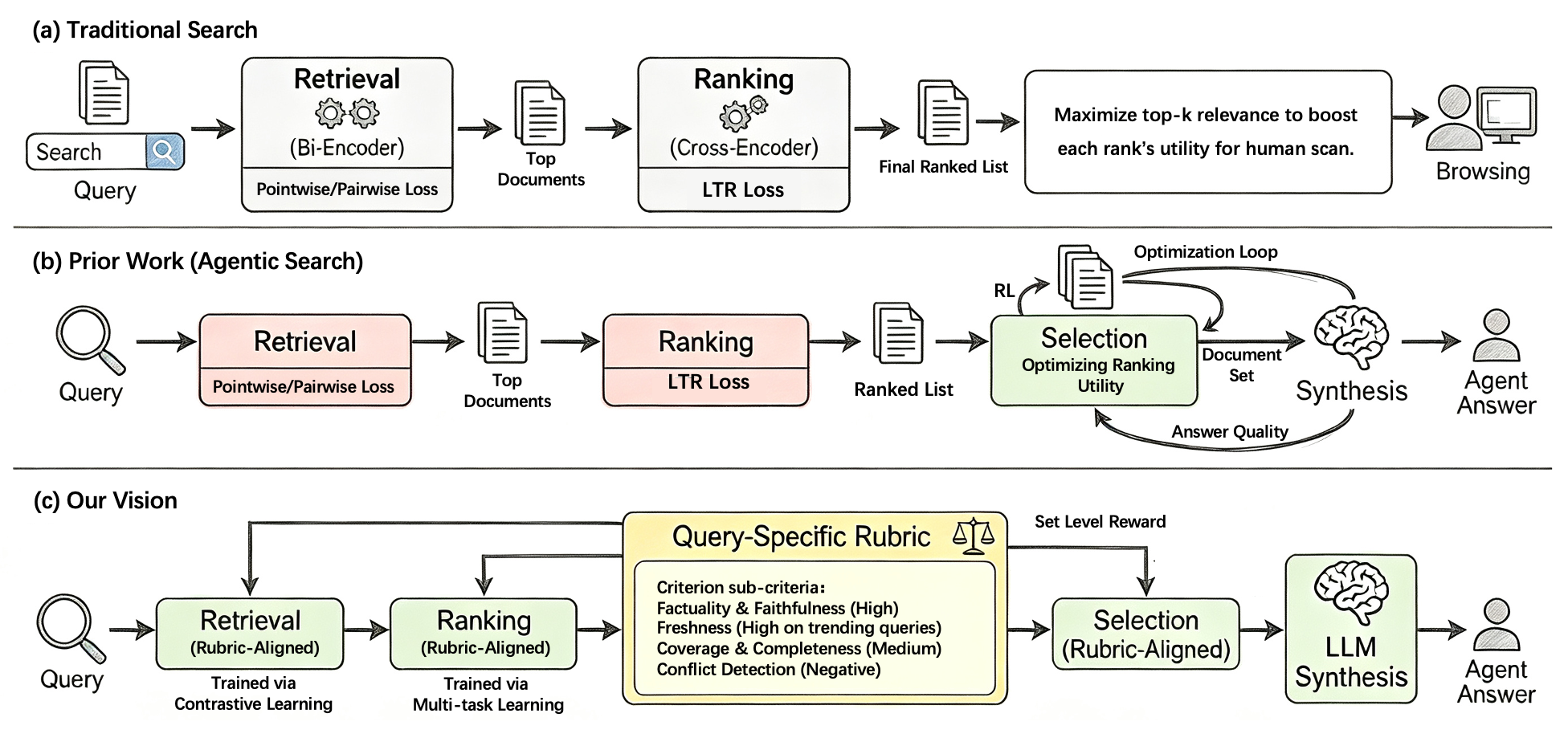}
\caption{%
(a) Traditional search pipelines score documents by individual relevance, optimized for human browsing over ranked lists.
(b) Prior agentic search work optimizes document selection for LLM synthesis, yet upstream retrieval and ranking remain unaligned to answer quality.
(c) Our proposed framework unifies the entire search pipeline under a query-specific rubric: the rubric provides per query--document supervision for retrieval and ranking, and set-level reward for document selection, aligning all stages toward high-quality answers within the same latency budget.}
\label{fig:motivation}
\end{figure}

\looseness=-1 To improve set-level utility, prior work has pursued two complementary
directions: characterizing evidence quality for LLM generation, and
optimizing evidence utility for downstream answer quality.
The first direction asks what makes a document set \emph{useful} for LLM
synthesis. Three aspects have been studied: task utility, whether the
evidence covers the information needed to answer the query
(e.g., answer contribution, information
coverage~\cite{wang2025infogainrag,lin2011submodular});
model usability, whether the model can effectively consume the
evidence (e.g., position bias, context
noise~\revfix{\cite{liu2023lostmiddle}}); and trustworthiness, whether the
evidence is faithful and free of hallucination (e.g., citation
faithfulness~\cite{gao2023alce}).
By decomposing holistic quality into discrete criteria, rubric-based
evaluators~\revfix{\cite{farzi2024rubric,hashemi2024rubric}}
can score document sets against all three aspects in one pass.
The second direction optimizes evidence utility through two lenses.
From the LLM training literature, RaR~\revfix{\cite{gunjal2025rar}} and
Chasing the Tail~\cite{zhang2026chasingtail} show that rubric-guided feedback serves as
an effective RL reward, yet neither has been applied to search
pipelines. From the retrieval literature, generator-aware selectors
such as SARA~\revfix{\cite{jin2025sara}} learn to choose document subsets that
maximize answer quality, but operate exclusively at the selection
stage, taking upstream retrieval and ranking as given.
However, prior work
either defines evidence quality without training it, or optimizes
evidence utility without aligning upstream stages
(Figure~\ref{fig:motivation}(b)). 

\looseness=-1 To close this gap, we adopt the rubric as the central optimization
signal for the entire search pipeline (Figure~\ref{fig:motivation}(c)). 
Translating rubric guidance into pipeline action, however, poses three non-trivial challenges.
First, there is a granularity mismatch between evaluation and training.
Retrieval and ranking stages must score documents independently to
enable efficient indexing and caching.  A rubric
evaluates document \emph{sets}, and its criteria cannot be applied to a
single document in isolation.
Second, annotation cost prohibits direct rubric-based labeling at scale.
Rubric-based evaluation demands costly LLM inference per query--document
set. Training retrieval and ranking models requires labeled data across
millions of query--document pairs, making per-instance annotation using strong reasoning LLMs
prohibitively expensive for production scale.
Third, even for the selection model, the only stage that natively
operates on sets, the rubric reward is inherently sparse: a single
scalar score assigned to an entire selected set provides no token-level
guidance, making credit assignment and training unstable.


To address these challenges, we first define a query-specific rubric with
positive criteria carrying importance weights (what the set must contain)
and negative criteria carrying deduction magnitudes (what it must avoid).
We design a coarse-to-fine procedure that
automates rubric construction at production scale. This formulation
reduces set-level evaluation to per-criterion scoring, making the rubric
actionable for training while the generation pipeline ensures rubrics are
both comprehensive and discriminative.
From this rubric we derive two training signals. The first is
per query--document utility labels. A large model first reasons with
the rubric to score individual query--document pairs. These scores
are distilled into a lightweight labeler via supervised fine-tuning
and refined through RL-based correction. The labeler thereby amortizes
rubric-grounded set-level evaluation into pointwise scores, resolving
the granularity mismatch noted above.
Guided by these utility labels, the retrieval stage trains a
contrastive dual-encoder (Qwen3-Embedding-0.6B~\cite{zhang2025qwen3embed}) with utility-aligned
fine-tuning, and the ranking stage trains a Qwen3-4B~\cite{qwen2025technical} ranker with a
utility-oriented focal loss.
The second signal is a set-level reward: a rubric-conditioned evaluator
scores candidate subsets directly against the rubric. We train the
selection stage via a dual-branch RL architecture: GRPO optimizes
the sparse set-level rubric reward, while Best-Rollout Distillation densifies credit assignment by treating the highest-reward
rollout as a token-level teacher.
We evaluate each component offline and in an online A/B test on Yuanbao
Search production traffic. Offline evaluation shows consistent gains
across pipeline stages, and in the downstream agent system, cumulative
answer quality improves by 8.52\%. Our contributions are:
\begin{enumerate}[leftmargin=*]
    \item A rubric-as-reward formulation that defines query-dependent,
    set-level utility as the single supervision surface for all pipeline
    stages, together with a lightweight labeler that amortizes this
    signal into per query--document utility labels at corpus scale;
    \item A dual-branch RL training strategy for set selection that combines GRPO with best-rollout distillation to ensure stable convergence under sparse set-level reward;
    \item An industry-deployed system in Yuanbao Search, with offline and
    online experiments validating the effectiveness of each pipeline
    component and the full agent search system.
\end{enumerate}

\section{Method}
\begin{figure*}[t]
\centering
\includegraphics[width=\textwidth]{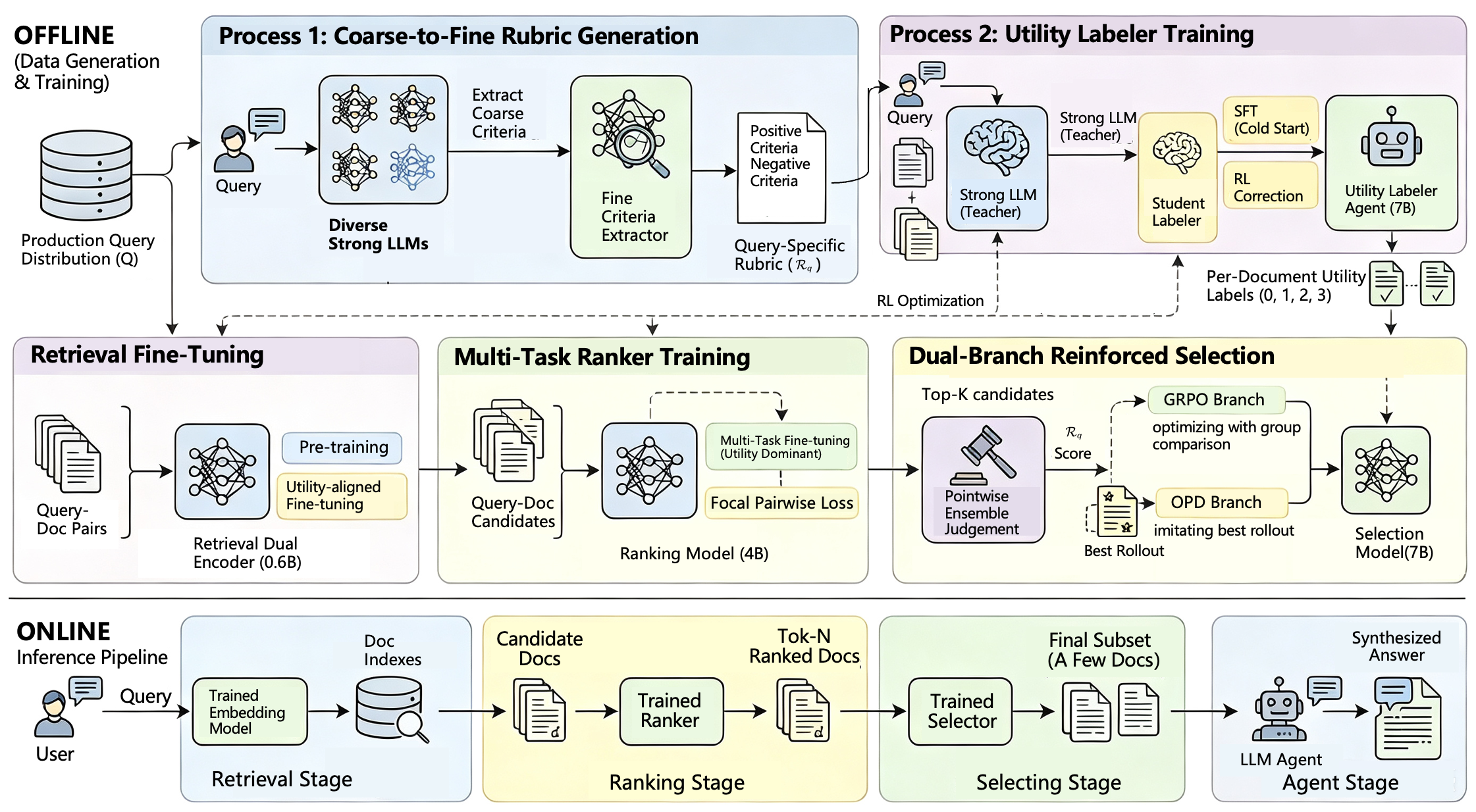}
\caption{Overall architecture of the rubric-guided agentic search framework.
\textbf{Offline (top)} involves coarse-to-fine rubric generation with
weighted positive and negative criteria, utility-labeler training by
teacher distillation, rubric-aligned retrieval and multi-task ranking,
and dual-branch selection with GRPO and best-rollout distillation (OPD).
\textbf{Online (bottom)} involves dense retrieval of candidate documents
with the trained models, rubric-aligned re-ranking, set selection, and
agent answer synthesis.}
\label{fig:architecture}
\end{figure*}

We address the aforementioned three challenges in three steps.
First, we define query-specific rubrics with weighted positive and
negative criteria, and automate their construction via a coarse-to-fine
generation pipeline.
Second, we train a labeler to amortize rubric-grounded evaluation into
per-document utility labels, supervising retrieval and ranking over the
full corpus.
Third, we train the selection stage via dual-branch RL with a
rubric-conditioned set-level reward, resolving credit assignment through
GRPO and best-rollout distillation.
Figure~\ref{fig:architecture} illustrates the overall framework.

\subsection{Rubric in Search}
\label{sec:rubric_in_search}
To operationalize utility, we adopt the concept of a \emph{rubric}, i.e., decomposing an information need
into discrete, verifiable criteria that collectively define what a dependable
evidence set looks like.

\subsubsection{Definition}
\label{sec:rubric_definition}
Let $q$ denote a user query drawn from the production query distribution
$\mathcal{Q}$ (i.e., $q \sim \mathcal{Q}$). Since an LLM agent needs different evidence per
query (e.g., timeliness for news, authority for medicine, completeness for
products), no static relevance schema suffices. 
We therefore associate each query $q$ with a \emph{rubric} $\mathcal{R}_q$.
Because not all criteria contribute equally to answer quality, each
carries an explicit weight that governs its contribution to the overall
score. Concretely, the rubric has two parts: 
(1) \textbf{Positive criteria} $\mathcal{R}_q^{+} = \{\, (p_i, c_i) \,\}_{i=1}^{I}$,
        where each $p_i$ is a required information facet and $c_i$ is its
        \emph{importance weight}, governing its proportion of the total score.
(2) \textbf{Negative criteria} $\mathcal{R}_q^{-} = \{\, (n_j, \eta_j) \,\}_{j=1}^{J}$,
        where each $n_j$ is a defect to avoid and $\eta_j$ is its
        \emph{deduction magnitude}, subtracted from the score when triggered.
A positive criterion is \emph{satisfied} when the document set
$\mathcal{D}_q$ collectively provides the required information; a
negative criterion is \emph{triggered} when any document in
$\mathcal{D}_q$ violates it. This makes scoring inherently set-level.
Positive and negative criteria define a utility
\emph{contract}: a document set is useful if it
satisfies weighted positive criteria and triggers no negative criteria.

\subsubsection{Coarse-to-Fine Rubric Generation}
\label{sec:rubric_generation}
To scale beyond a seed set of manually authored rubrics, we decompose rubric generation
into two stages. The coarse stage ensures comprehensiveness; the fine stage ensures
specificity.

\noindent\textbf{Coarse stage.}
Given $q$, three diverse LLMs (ChatGPT, Claude, and GLM) independently
generate free-form answers\revfix{, each grounded in live web retrieval over
multiple external search engines}, from which an extraction prompt
$P_{\mathrm{gen}}$ surfaces candidate criteria
$\mathcal{R}_q^{\mathrm{cand},(k)}$ for each LLM $k\in\{1,2,3\}$,
covering required information facets. Extracting criteria from answers
rather than directly from $q$ leverages the natural information
faceting that answer generation provides; aggregating across models
captures a broader range of criteria than any single model would. A merge prompt
$P_{\mathrm{merge}}$ aligns, deduplicates, and structures the
three sets into a coarse skeleton $\mathcal{R}_q^{\mathrm{coarse}}$
of high-level dimensions (e.g., ``Performance,'' ``Pricing'')
that capture what matters but lack per-document specificity.

\noindent\textbf{Fine stage.}
We anchor $\mathcal{R}_q^{\mathrm{coarse}}$ to the top-$M$ documents
$\mathcal{D}_{\mathrm{top}}$ retrieved by the production system
($M = 15$), since a rubric must
distinguish documents in the actual candidate pool rather than in the abstract.
Given $q$, $\mathcal{D}_{\mathrm{top}}$, and
$\mathcal{R}_q^{\mathrm{coarse}}$, an LLM extracts specific, verifiable sub-criteria, producing
the final rubric $\mathcal{R}_q$ with both weighted positive criteria and negative criteria carrying deduction magnitudes (e.g., ``Performance''
$\rightarrow$ ``3.5$\times$ AI performance over the prior generation''). Each
sub-criterion passes through a fact-verification gate: claims involving
numerical values, entities, or time-sensitive facts are validated
against search-engine evidence; logical constraints pass directly.


Given $\mathcal{R}_q$, we then convert it into training signals
through two paths. First, a labeler consumes
$\mathcal{R}_q$ and emits per-document utility labels, which supervise
retrieval and ranking over
the full corpus. Second, a pointwise-ensemble judge
 scores subsets directly against $\mathcal{R}_q$,
producing the set-level reward that drives selection training.

\subsection{Rubric-oriented Utility Label Generation}
\label{sec:labeler}
\subsubsection{Utility Label Generation}
The traditional retrieval--ranking funnel scores documents independently
to support efficient indexing and top-$K$ retrieval, and therefore
requires pointwise supervision. We define a labeler $L_{\theta}:
(q, d, \mathcal{R}_q) \mapsto l \in \{0, 1, 2, 3\}$ that approximates
set-level rubric satisfaction with a per-document grade.
A document receives $l = 3$ when it covers all high-weight $p_i
\in \mathcal{R}_q^{+}$ and triggers no $n_j \in \mathcal{R}_q^{-}$
(complete evidence).
$l = 2$ indicates that most high-weight criteria are covered but
minor ones are missed (sufficient evidence).
Documents covering only low-weight criteria earn $l = 1$ (auxiliary
context).
$l = 0$ marks documents that trigger any $n_j$ or cover no $p_i$
(useless or harmful).
For example, given the query ``2026 iPad Pro performance and price'', a document with M5
benchmarks, battery, and official pricing receives $l = 3$; one
mentioning only the chip name receives $l = 1$.
For simple queries $l{=}3$ is attainable by a single document; for complex queries with many facets, even a partial match at $l{=}2$ reduces the documents needed to satisfy $\mathcal{R}_q$, so higher grades remain informative.

Assigning these labels requires multi-criterion reasoning: the model must
check each $(q, d)$ pair against every positive and negative criterion in
$\mathcal{R}_q$, weigh their relative importance, and produce both a label
and a rationale explaining which criteria are met or violated. A small
model lacks the reasoning capacity to do this reliably across diverse
queries, while running a strong LLM on every query-document pair is
infeasible at corpus scale.
We resolve this tension through distillation. A strong LLM serves as an
offline teacher, annotating a small subset of query-document pairs with
a utility grade $\hat{l}$ and a rationale $\hat{\mathbf{e}}$.
A lightweight student labeler is then trained to
approximate the teacher's judgment and scales annotation to millions of
query-document pairs at negligible cost.

	\subsubsection{Labeler Training}
	\label{sec:labeler_training}
	We train the lightweight labeler in two stages.
	In the SFT stage, the student is fine-tuned on teacher-annotated
	$(q, d, \mathcal{R}_q) \rightarrow (\hat{l}, \hat{\mathbf{e}})$ pairs
	via next-token prediction, learning to follow the labeling format
	and produce grounded rationales. This provides a necessary cold start:
	without it, RL would explore an unstructured output space with no
	signal. However, SFT optimizes for surface imitation rather than
	rubric satisfaction, so its accuracy is bounded by the teacher.
	We address this gap with an RL stage that directly optimizes rubric alignment.
		The reward combines three terms, each targeting a distinct
failure mode:
\begin{equation}
R_{\mathrm{label}} = \lambda_{\mathrm{acc}} R_{\mathrm{acc}} + \lambda_{\mathrm{rank}} R_{\mathrm{rank}} + \lambda_{\mathrm{fmt}} R_{\mathrm{fmt}},
\label{eq:labeler_reward}
\end{equation}
where $R_{\mathrm{acc}}$ rewards exact-match accuracy against
rubric-verified labels, $R_{\mathrm{rank}}$ rewards pairwise order
correctness over a defined subset of documents, and $R_{\mathrm{fmt}}$ enforces valid output structure.
Together, SFT and RL produce a labeler that approaches the teacher's
quality at a fraction of the inference cost.

\subsection{Rubric-Aligned Retrieval}
\label{sec:retrieval}

Given $q$, the retrieval stage returns a candidate set $\mathcal{D}_q$ of roughly one thousand documents from
a billion-scale corpus. We use a lightweight
dual encoder (Qwen3-Embedding-0.6B~\cite{zhang2025qwen3embed}), trained in two stages.

\subsubsection{In-domain contrastive pre-training.}

We pre-train on 610M production $(q, d)$ pairs. For each query,
the logs provide $K$ co-occurring documents; one is sampled as the pseudo-positive
$d^{+}$, and the remaining $K-1$ serve as group-level hard negatives
$\mathcal{N}_q^{\mathrm{group}}$. These share topical overlap with $q$, making
them harder than batch-level negatives $\mathcal{N}_q^{\mathrm{batch}}$ drawn from
other queries. The encoder is optimized with bidirectional InfoNCE:
\begin{equation}
\small
\mathcal{L}_{q \rightarrow d} = -\log
\frac{\exp(s(q, d^+)/\tau)}
{\exp(s(q, d^+)/\tau) + \sum_{d^- \in \mathcal{N}_q^{-}} \exp(s(q, d^-)/\tau)},
\label{eq:retrieval_group}
\end{equation}
where $\mathcal{N}_q^{-} = \mathcal{N}_q^{\mathrm{group}} \cup
\mathcal{N}_q^{\mathrm{batch}}$, $s(\cdot,\cdot)$ is cosine similarity, and
$\tau$ is a temperature. We optimize the symmetric loss
$\mathcal{L}_{\mathrm{pretrain}} = \mathcal{L}_{q \rightarrow d} +
\mathcal{L}_{d \rightarrow q}$. Matryoshka Representation Learning
(MRL)~\cite{kusupati2022mrl} is applied so a single checkpoint serves multiple
embedding dimensionalities.

\subsubsection{Utility-aligned fine-tuning.}

Pre-training uses query-log co-occurrence as supervision, which does not
distinguish a document that fully satisfies $\mathcal{R}_q$ from one covering
only a minor facet. We fine-tune using the labeler's graded labels
: documents with $l \geq 2$ serve as positives, and those
with $l < 2$ form group-level negatives. The loss is identical to
Eq.~\ref{eq:retrieval_group}. When multiple documents per query satisfy
$l \geq 2$, we sample one as $d^{+}$ and mask the remaining positives from batch
negatives to avoid penalizing them. This stage aligns the encoder's
representation space with rubric-grounded utility.

\subsection{Multi-Task Ranking}
\label{sec:ranking}

This section describes the architecture and training objectives of our utility-aligned ranking model.
The ranking stage reorders $\mathcal{D}_q$ into a ranked list
$\mathcal{D}_{\mathrm{rank}}$ of roughly twenty documents for selection
. We use a Qwen3-4B~\cite{qwen2025technical}
decoder-only ranker, trained with four scoring heads sharing a single
encoder: utility, authority, freshness, and quality. The utility head is
supervised by the labeler's rubric-aligned annotations
; the other three are supervised by existing
production signals. At inference, only the utility head is used,
producing a single scalar ranking score.
The auxiliary heads serve as multi-task regularizers~\cite{gu2020dmt} that encourage the shared encoder to learn authority-, freshness-, and quality-aware representations, which indirectly benefit utility scoring.
Each head is optimized with a pairwise loss. We down-weight the three
auxiliary losses by $\lambda_{\mathrm{aux}} = 0.5$, giving the utility
objective twice the per-head gradient weight of any single
auxiliary task.
This ensures utility scoring remains the primary training signal while
authority, freshness, and quality provide moderate regularization:
\begin{equation}
\small
\mathcal{L}_{\mathrm{rank}} = \mathcal{L}_{\mathrm{utility}}
    + \lambda_{\mathrm{aux}} \left( \mathcal{L}_{\mathrm{authority}}
    + \mathcal{L}_{\mathrm{freshness}}
    + \mathcal{L}_{\mathrm{quality}} \right).
\label{eq:rank_loss}
\end{equation}
In late-stage training, easy pairs dominate the gradient. We apply focal
reweighting~\cite{lin2017focal} to each head's pairwise loss, upweighting pairs where the
student score gap lags behind the teacher label gap:
\begin{equation}
\small
\mathcal{L}_{\mathrm{dim}} = \frac{1}{|\mathcal{P}|}
    \sum_{(i,j) \in \mathcal{P}}
    (1 - \hat{p}_{ij})^\gamma \cdot
    \mathrm{ReLU}\bigl(\Delta y_{ij} - \Delta s_{ij}\bigr),
\label{eq:focal_loss}
\end{equation}
where $\hat{p}_{ij} = \sigma(s_i - s_j)$, $\Delta y_{ij}$ is the
label gap, $\Delta s_{ij}$ is the predicted score gap, and $\gamma = 2.0$.
Finally, we fine-tune on \revfix{human-annotated samples} with a reduced learning rate to sharpen alignment with rubric-conditioned utility labels.

\subsection{Dual-Branch Reinforced Selection}
\label{sec:selection}

The selection stage takes the ranked list $\mathcal{D}_{\mathrm{rank}}$
(roughly twenty documents) and selects a final subset $\mathcal{S}_q^*$
of 3--8 documents. Unlike retrieval and ranking, which score documents
independently, selection optimizes subsets collectively for rubric
coverage under diversity constraints. We train a
Qwen3-8B~\cite{qwen2025technical} selector as a policy
$\pi_\theta$ via reinforcement learning, with rewards provided by a
Pointwise-Ensemble (PE) Judge $\mathcal{J}$ (\revfix{DeepSeek-V4-Flash}) that
scores candidate subsets $\mathcal{S}$ directly against $\mathcal{R}_q$
by aggregating weighted criterion satisfaction minus triggered deductions
(\S\ref{sec:rubric_definition}). The core challenge is credit
assignment: the reward $r(q, \mathcal{S})$ is a single scalar per
subset, with no token-level signal. We address this through
dual-branch training with group-relative advantages.

For each query, we sample $G = 8$ rollouts $\{\mathcal{S}^{(g)}\}_{g=1}^{G}$
from $\pi_\theta$ given $\mathcal{D}_{\mathrm{rank}}$, score them with
$\mathcal{J}$ to obtain $r^{(g)} = \mathcal{J}(q, \mathcal{S}^{(g)}, \mathcal{R}_q)$,
and compute group-normalized advantages
$A^{(g)} = (r^{(g)} - \mu) / (\sigma + \varepsilon)$, where $\mu, \sigma$
are the group mean and standard deviation of $\{r^{(g)}\}$ and
$\varepsilon = 10^{-4}$.
Rollouts with $r^{(g)} > \mu$ receive positive advantages and enter the
\textbf{GRPO branch}, where a clipped surrogate objective reinforces
successful patterns at the token level:
\begin{equation}
\footnotesize
\mathcal{L}_{\mathrm{GRPO}} = -\mathbb{E}_{g: r^{(g)} > \mu}
\Bigg[ \frac{1}{T^{(g)}} \sum_{t=1}^{T^{(g)}}
\min\Big(\rho_t^{(g)} A^{(g)},\;
\mathrm{clip}(\rho_t^{(g)}, 1-\epsilon, 1+\epsilon)\, A^{(g)}\Big) \Bigg],
\label{eq:grpo}
\end{equation}
where $\rho_t^{(g)} = \pi_\theta(s_t^{(g)} \mid q, \mathcal{D}_{\mathrm{rank}}, s_{<t}^{(g)}) /
\pi_{\theta_{\mathrm{old}}}(s_t^{(g)} \mid q, \mathcal{D}_{\mathrm{rank}}, s_{<t}^{(g)})$
and $\epsilon = 0.2$. A KL penalty against a reference model prevents
policy degradation.
For rollouts with $r^{(g)} \leq \mu$, negative advantages would
indiscriminately suppress all token choices. We instead distill the
best in-group rollout $\mathcal{S}^\star = \arg\max_g r^{(g)}$ into
the policy via weighted likelihood maximization, with correction
strength proportional to the reward gap $w_g = r^\star - r^{(g)}$
where $r^\star = \max_g r^{(g)}$:
\begin{equation}
\mathcal{L}_{\mathrm{distill}} = -\mathbb{E}_{g: r^{(g)} \leq \mu}
\left[ \frac{w_g}{T^\star} \sum_{t=1}^{T^\star} \log \pi_\theta(s_t^\star \mid q, \mathcal{D}_{\mathrm{rank}}, s_{<t}^\star) \right].
\label{eq:distill}
\end{equation}
The joint loss $\mathcal{L}_{\mathrm{sel}} = \mathcal{L}_{\mathrm{GRPO}} + \lambda_{\mathrm{distill}} \mathcal{L}_{\mathrm{distill}}$
($\lambda_{\mathrm{distill}} = 0.5$) decomposes credit assignment into
complementary signals: GRPO reinforces above-average patterns via
group-relative advantage, while distillation corrects below-average
rollouts by pulling them toward the best in-group solution.
$\mathcal{S}_q^*$ feeds the agent for answer generation.

\section{Experiments}
To validate the effectiveness of the proposed rubric-aligned pipeline, we conduct extensive offline and online experiments on Yuanbao Search, a large-scale commercial AI search engine.
\subsection{Experimental Setup}
\label{sec:experimental_setup}

\subsubsection{Dataset}
\label{sec:dataset}
\looseness=-1 We train and evaluate our pipeline using both logged behavioral data and human-annotated data.
\revfix{Training uses 62M production queries with 610M query-document pairs for rubric-aligned retrieval and ranking, and 180K queries with 3.2M pairs for selection training.}
Human-annotated data serves supervised fine-tuning and evaluation for each downstream stage.
For \textbf{retrieval} evaluation, we construct four held-out benchmarks covering diverse query distributions:
\textbf{Random-1K} (1,000 randomly sampled queries, \revfix{3.8M pairs}),
\textbf{Long-Tail} (449 queries with \revfix{fewer than 2.5\% relevant documents}, \revfix{1.6M pairs}),
\textbf{Entity} (707 queries containing named entities such as persons, organizations, and locations, \revfix{2.7M pairs}), and
\textbf{English} (276 English queries, \revfix{780K pairs}).
Each benchmark pairs queries with candidate documents from the production first-stage retriever; ground-truth relevance is assigned by human annotators guided by the rubric.
For \textbf{labeling}, \revfix{we annotate 200K query-document pairs drawn from 28K production queries.} Each pair is annotated by a strong LLM teacher given the query-specific rubric $\mathcal{R}_q$, producing utility grades $l \in \{0,1,2,3\}$.
For \textbf{ranking} and \textbf{selection} offline evaluation, we sample 400,000 queries (\revfix{20M query-document pairs}, 50 documents per query) from online production traffic with candidate documents from the upstream retriever. The query distribution matches production traffic: informational (35\%), hard queries with key-term omission and content mismatch (25\%), sensitive queries on authority/timeliness/medical domains (35\%), and English (5\%). Pairwise relevance judgments are derived from human evaluators using satisfaction-oriented labels, from which we compute PNR as the primary metric.
\looseness=-1 For \textbf{end-to-end} online evaluation, we conduct A/B testing on 5\% of production traffic over 14 days. From the 14-day traffic, 1,000 queries are sampled for blind side-by-side human assessment, where annotators compare treatment vs.\ baseline answers without knowing which system produced each response.
Table~\ref{tab:dataset} summarizes the data.

\begin{table}[t]
\small
\centering
\caption{Dataset statistics across pipeline stages.}
\label{tab:dataset}
\begin{tabular}{l r r}
\toprule
\textbf{Stage} & \textbf{\#Queries} & \textbf{\#Query-Doc Pairs} \\
\midrule
Training corpus                   & 62,000,000   & 610,000,000 \\
Retrieval (Random-1K)            & 1,000        & 3,800,000 \\
Retrieval (Long-Tail)            & 449          & 1,600,000 \\
Retrieval (Entity)               & 707          & 2,700,000 \\
Retrieval (English)              & 276          & 780,000 \\
Labeling                         & 28,000       & 200,000 \\
Ranking \& Selection (offline)   & 400,000      & \revfix{20,000,000} \\
Selection (training)             & 180,000      & 3,200,000 \\
\bottomrule
\end{tabular}
\end{table}

\subsubsection{Evaluation Methodology}
\label{sec:evaluation_methodology}

We evaluate each pipeline stage with dedicated metrics: Recall@1K for retrieval; PNR for labeling and ranking; per-document rubric scores and the PE Judge set-level score for selection; GSB for end-to-end answer quality.

\paragraph{Positive-Negative Ratio (PNR)}
PNR measures pairwise ranking alignment: given query $q$, $N$ documents with labels $y_i$ and predicted scores $f(q,d_i)$,
\begin{equation}
    \mathrm{PNR} = \frac{\sum_{i,j \in [1,N]} \mathbf{1}[y_i > y_j] \cdot \mathbf{1}[f(q,d_i) > f(q,d_j)]}
                       {\sum_{i,j \in [1,N]} \mathbf{1}[y_i > y_j] \cdot \mathbf{1}[f(q,d_i) < f(q,d_j)]},
\end{equation}
where $\mathbf{1}[\cdot]$ is the indicator function. PNR $> 1$ indicates correct pairwise ordering; higher is better. We report micro-averaged PNR with $\epsilon = 0.1$ for numerical stability.

\paragraph{Good-Same-Bad (GSB)}
GSB measures end-to-end answer quality via blind side-by-side human evaluation. Annotators judge whether the treatment response is better (Good), worse (Bad), or comparable (Same) against a baseline. The aggregated lift is:
\begin{equation}
    \Delta\mathrm{GSB} = \frac{N_{\mathrm{Good}} - N_{\mathrm{Bad}}}{N_{\mathrm{Good}} + N_{\mathrm{Same}} + N_{\mathrm{Bad}}}.
\end{equation}
Annotators follow the labeling pipeline's calibration protocol.

\subsubsection{Competitor Systems}
\label{sec:competitor_systems}

Recent agentic-search methods~\revfix{\cite{li-etal-2025-search,curiosity2026}}
and RAG subset selection~\revfix{\cite{contextpicker2025,jin2025sara,fan2026rank4gen}}
achieve strong results, but are typically evaluated on corpora of
$10$--$10^2$ documents with relaxed latency.
Production search engines operate over $10^3$-scale candidate pools
under \revfix{a fixed serving-latency budget (hundreds of milliseconds per stage)}, so scaling evidence selection to
thousands of candidates is a system-level problem. We evaluate each stage
against production baselines of comparable scale.
\begin{itemize}[leftmargin=*]
  \item \textbf{Retrieval-Base (Zero-shot)}: Qwen3-Embedding-0.6B~\cite{zhang2025qwen3embed}
        zero-shot checkpoint, measuring out-of-the-box semantic matching.
  \item \textbf{Retrieval-Base (Instruct)}: Qwen3-Embedding-0.6B-Instruct,
        the instruction-tuned variant~\cite{zhang2025qwen3embed} with
task-specific prefixes.
  \item \textbf{Ranking-Base}: The production ranker on Yuanbao Search:
        a Qwen3-4B model trained with pairwise ranking on large-scale
        click-log relevance judgments~\cite{ma2024rankllama}, with click
        modeling, position-bias correction, and freshness features.
  \item \textbf{Selector-Base}: The production selection policy on Yuanbao
        Search: a Qwen3-8B model predicting per-document relevance and
        authority from production feedback, selecting top-$k$ documents
        with submodular diversity
        re-ranking~\cite{lin2011submodular,carbonell1998mmr}, incorporating
        engagement signals (clicks, dwell time, task completion) and
        explicit diversity constraints.
\end{itemize}

\paragraph{Implementation.}\tolerance=800

For the \textbf{labeler}, a \revfix{Qwen3-4B} student is distilled from a Qwen3-235B
teacher on \revfix{a broad set of query-document pairs}. SFT uses
lr~\mbox{$2\times10^{-5}$}, batch size 512, max length 4096, 3 epochs, gradient
clipping 1.0. RL uses GRPO with lr~\mbox{$5\times10^{-7}$}, 8 rollouts per query,
KL penalty $\beta_{\mathrm{KL}}{=}0.01$, reward weights $\lambda_{\mathrm{acc}}{=}0.5$, $\lambda_{\mathrm{rank}}{=}0.3$, $\lambda_{\mathrm{fmt}}{=}0.2$.
For \textbf{retrieval}, a Qwen3-Embedding-0.6B is trained in two stages on
610M in-domain $(q, d)$ pairs. Stage~1 (in-domain contrastive
pre-training) uses bidirectional InfoNCE ($\tau{=}0.05$, $B{=}64$ queries per batch),
lr~\mbox{$5\times10^{-5}$}, 3 epochs, max length
512, and MRL across $\{128, 256, 512, 768\}$ (final dimension 768).
Stage~2 (utility-aligned fine-tuning) uses the same loss with
labeler-identified positives ($l{\geq}2$), lr~\mbox{$2\times10^{-5}$}, batch size 512,
1 epoch; served with product quantization.
For \textbf{ranking}, a Qwen3-4B is trained on production query-document
pairs labeled by the labeler, with focal pairwise loss ($\gamma{=}2.0$) over
four scoring heads (utility, authority, freshness, quality), auxiliary loss weight $\lambda_{\mathrm{aux}}{=}0.5$, lr~\mbox{$1\times10^{-5}$},
batch size 256, 2 epochs, max length 4096, pairs constructed within-query.
\revfix{Human-annotated fine-tuning uses a reduced learning rate.}
For \textbf{selection}, a Qwen3-8B is \revfix{warm-started by distilling a
strong teacher on $\sim$300K samples}, then trained via dual-branch RL on
180K training queries, with $G{=}8$ rollouts per query, lr~\mbox{$5\times10^{-7}$},
KL penalty $\beta_{\mathrm{KL}}{=}0.01$, rollout temperature 1.0, max generation length
2048, GRPO clip $\epsilon{=}0.2$, distillation weight $\lambda_{\mathrm{distill}}{=}0.5$. The PE Judge
uses \revfix{DeepSeek-V4-Flash} to score subsets against $\mathcal{R}_q$.
All models are trained using AdamW
($\beta_1{=}0.9$, $\beta_2{=}0.95$, weight decay $0.01$) with 10\% linear
warmup and cosine decay schedule. \revfix{The full offline pipeline, including rubric
generation, labeler training, and selector training, takes 1--2 weeks of wall-clock training time.}

\subsection{Online Evaluation}
\label{sec:experiments_overall}
We deploy each pipeline component independently on Yuanbao Search, with A/B testing on 5\% of production traffic over 14 days. Table~\ref{tab:incremental} reports the GSB lift of each component independently added to the production baseline. All three components yield significant gains ($p<0.05$).
Selection provides the largest gain ($+4.82\%$): set-level optimization directly targets final subset quality, which GSB is designed to measure. Ranking adds $+3.94\%$ through rubric-aware multi-dimensional scoring and backbone scaling. Retrieval contributes $+0.40\%$; while modest, its impact is cost-effective since the Qwen3-Embedding-0.6B adds negligible inference cost and its rubric-aligned training propagates higher-quality candidates to all downstream stages.
The sum of individual lifts ($+9.16\%$) exceeds the end-to-end net gain ($+8.52\%$) by $+0.64$ percentage points. This gap reflects overlap among the three components, which are trained toward the same rubric-aligned objective so their benefits are not strictly additive. \textbf{The $+8.52\%$ net gain is the actual improvement experienced by end users.}

\begin{table}[t]
\centering
\caption{Incremental online A/B results (agent answer GSB lift).
Each row adds the indicated component independently to the production baseline.}
\label{tab:incremental}
\small
\begin{tabular}{lc}
\toprule
\textbf{Configuration} & $\Delta$\textbf{GSB (\%)} \\
\midrule
Baseline (production)                       & --- \\
\quad +\ Rubric-Aligned Retrieval           & +0.40$^\ast$ \\
\quad +\ Rubric-Aligned Ranking             & +3.94$^\ast$ \\
\quad +\ Rubric-Aligned Selection           & +4.82$^\ast$ \\
\midrule
Sum of individual lifts                     & +9.16 \\
Full pipeline (all components)              & +8.52 \\
\bottomrule
\multicolumn{2}{p{0.92\columnwidth}}{\small\vspace{2pt}
$^\ast$: significant over the production baseline
(paired $t$-test, $p<0.05$).} \\
\end{tabular}
\end{table}

\subsection{Rubric Quality Analysis}
\label{sec:experiments_rubric}

\revfix{To evaluate the effectiveness of our rubrics, we compare model scoring
with and without rubrics against human judgment on a held-out,
human-annotated set of 2,000 pairwise set-ranking samples. Each sample is a
query, two candidate sets $A$ and $B$, and the human verdict on which set is
better, judged on utility, time-sensitivity, authority need, and freshness
need. We score both sets with the reward model under the training settings and
compare its verdict with the human verdict, so every number below is a
pairwise agreement rate. Table~\ref{tab:rubric_ablation} reports the results.}

\begin{table}[h]
\centering
\caption{\revfix{Pairwise agreement between the reward model's verdict and human
verdicts on 2,000 set-ranking pairs. The first two rows compare scoring with and
without a rubric. The remaining rows remove one rubric component at a time from
the full rubric.}}
\label{tab:rubric_ablation}
\small
\setlength{\tabcolsep}{6pt}
\begin{tabular}{@{}l c@{}}
\toprule
\textbf{Setting} & \textbf{Agreement} \\
\midrule
LLM scores utility directly (no rubric) & 68.7\% \\
\textbf{Full rubric (ours)} & \textbf{79.3\%} \\
\midrule
w/o criterion weights (uniform) & 79.0\% \\
w/o multi-model generation (single LLM) & 77.7\% \\
w/o fine sub-criteria (no top-15 anchoring) & 74.3\% \\
w/o negative criteria (positives only) & 72.8\% \\
\bottomrule
\end{tabular}
\end{table}

\revfix{Removing the rubric drops agreement from 79.3\% to 68.7\%, and removing
any single component also lowers it: negative criteria ($-6.5$), fine
sub-criteria ($-5.0$), multi-model generation ($-1.6$), and criterion weights
($-0.3$). Each component is therefore effective. Agreement tops out at
79.3\%; of the remaining 21\%, 7\% are genuine misses, 4\% are queries with
unclear intent where neither set is clearly better, and 10\% are
partial-satisfaction cases.}

\subsection{Labeling Quality Analysis}
\label{sec:experiments_labeler}

The labeler $L_\theta$ maps each
$(q, d, \mathcal{R}_q)$ to a utility grade $l \in \{0,1,2,3\}$. A \revfix{Qwen3-4B} backbone is warm-started via SFT on
\revfix{teacher-annotated $(q,d)$ pairs}, then refined via GRPO-based RL with the
hybrid reward.
Table~\ref{tab:labeler_offline} reports utility-oriented PNR (4-class) on an
internal benchmark. The \textbf{SFT baseline} uses the \revfix{base annotations}.
\textbf{+Data optimization} augments $\mathcal{D}_{\mathrm{ann}}$ with hard cases
annotated via chain-of-thought prompting,
which improves labeling coverage at the cost of increased inference overhead.
\textbf{+RL optimization} further fine-tunes the SFT checkpoint via GRPO with
the rubric-aligned hybrid reward.

\begin{table}[t]
\centering
\caption{Labeler PNR under successive training strategies.}
\label{tab:labeler_offline}
\begin{tabular}{l c}
\toprule
\textbf{Configuration} & \textbf{PNR} \\
\midrule
SFT baseline                         & 4.50 \\
\ +\ Data optimization               & 4.67$^\ast$ \\
\textbf{+\ RL optimization}          & \textbf{4.82}$^\ast$ \\
\bottomrule
\multicolumn{2}{c}{$^\ast$: significant over the SFT baseline (paired $t$-test, $p<0.05$).} \\
\end{tabular}
\end{table}
We make two observations. \textbf{First, data quality alone lifts PNR by $+0.17$ over the SFT baseline} ($p<0.05$, paired $t$-test): adding targeted hard cases and chain-of-thought prompting improves labeling coverage, confirming that rubric-aligned annotation quality depends critically on training data curation.
Second, \textbf{RL optimization adds a further $+0.15$ PNR} ($p<0.05$ vs.\ data optimization), comparable in magnitude to the data gain, confirming that SFT provides a necessary cold start but directly optimizing rubric alignment through RL is essential to surpass the teacher's ceiling.
The RL-trained checkpoint (PNR~4.82) is deployed as the production labeler, providing supervision for retrieval and ranking.





\subsection{Retrieval and Ranking Analysis}
\label{sec:experiments_retrieval_ranking}

\subsubsection{Retrieval Component Ablation}

We ablate each training component starting from Retrieval-Base, the open-source
Qwen3-Embedding\-0.6B zero-shot checkpoint. Components are added
incrementally: in-domain contrastive pre-training on 610M in-domain
$(q, d)$ pairs with bidirectional InfoNCE (+\ In-domain pretrain),
followed by utility-aligned fine-tuning where labeler-verified positives ($l{\geq}2$) replace query-log
positives in the same loss
(+\ Utility-aligned FT).

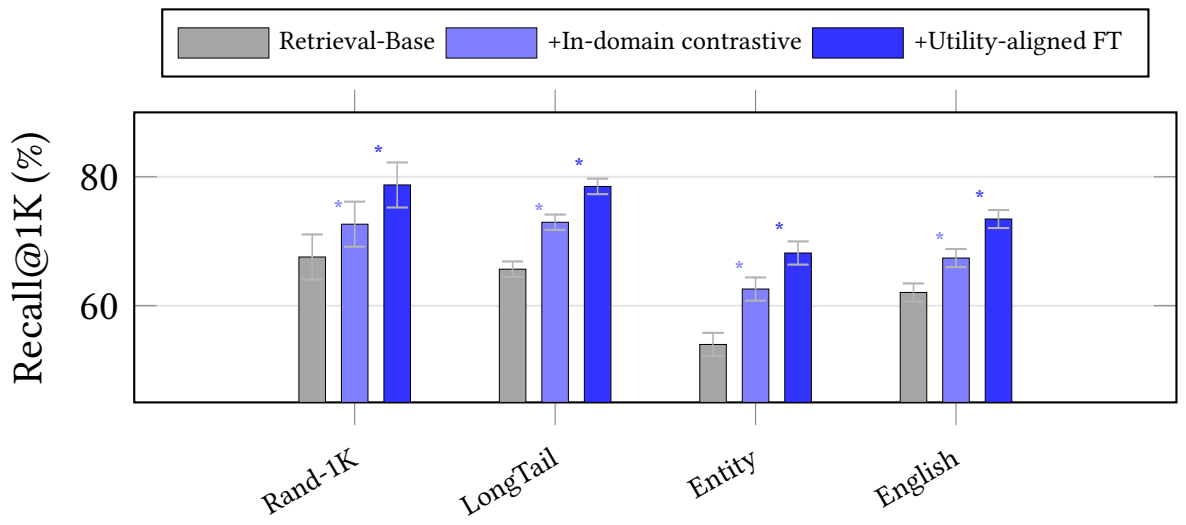
\begin{figure}[t]
\centering
\small
\begin{tikzpicture}[
    bar/.style={line width=0pt},
]
\begin{axis}[
    width=\columnwidth,
    height=3.5cm,
    ybar=3pt,
    bar width=5pt,
    ymin=45, ymax=90,
    ylabel={Recall@1K (\%)},
    xtick=data,
    xmin=0.5, xmax=4.5,
    xticklabels={Rand-1K, LongTail, Entity, English},
    xticklabel style={rotate=30, anchor=north east, font=\fontsize{5.5}{6.5}\selectfont},
    legend style={font=\fontsize{5}{6}\selectfont, legend columns=3,
                  at={(0.5,1.12)}, anchor=south},
    ymajorgrids=true, grid style=gray!20,
    enlarge x limits=0.15,
]
    \addplot[bar, fill=black!35, area legend,
             error bars/.cd, y dir=both, y explicit, error bar style={gray!60, line width=0.3pt}]
             coordinates {(1,67.56) +- (0,3.5) (2,65.67) +- (0,1.2) (3,53.99) +- (0,1.8) (4,62.06) +- (0,1.4)};
    \addplot[bar, fill=blue!50,  area legend,
             error bars/.cd, y dir=both, y explicit, error bar style={gray!60, line width=0.3pt}]
             coordinates {(1,72.65) +- (0,3.5) (2,72.95) +- (0,1.2) (3,62.59) +- (0,1.8) (4,67.40) +- (0,1.4)};
    \addplot[bar, fill=blue!80,  area legend,
             error bars/.cd, y dir=both, y explicit, error bar style={gray!60, line width=0.3pt}]
             coordinates {(1,78.74) +- (0,3.5) (2,78.51) +- (0,1.2) (3,68.18) +- (0,1.8) (4,73.45) +- (0,1.4)};
    \legend{Retrieval-Base, +In-domain contrastive, +Utility-aligned FT}
    \node[font=\tiny\bfseries, text=blue!80] at (axis cs:1.12, 83.5) {*};
    \node[font=\tiny\bfseries, text=blue!80] at (axis cs:2.12, 81.5) {*};
    \node[font=\tiny\bfseries, text=blue!80] at (axis cs:3.12, 71.5) {*};
    \node[font=\tiny\bfseries, text=blue!80] at (axis cs:4.12, 76.5) {*};
    \node[font=\tiny\bfseries, text=blue!50] at (axis cs:0.92, 75.0) {*};
    \node[font=\tiny\bfseries, text=blue!50] at (axis cs:1.92, 74.5) {*};
    \node[font=\tiny\bfseries, text=blue!50] at (axis cs:2.92, 65.5) {*};
    \node[font=\tiny\bfseries, text=blue!50] at (axis cs:3.92, 70.0) {*};
\end{axis}
\end{tikzpicture}
\caption{Component ablation of rubric-aligned retrieval. Recall@1K on four
internal benchmarks. Error bars denote $\pm 1$ standard error over queries.
$^\ast$: significant over the previous step
(paired $t$-test, $p<0.05$).}
\label{fig:retrieval_main}
\end{figure}

In-domain pre-training contributes the largest gain
($+5.1$--$+8.6$ points over Retrieval-Base), confirming that domain
adaptation is critical. Utility-aligned fine-tuning ($l{\geq}2$) adds a
further $+5.6$--$+6.1$ points by replacing query-log co-occurrence signals
with rubric-grounded supervision. The combined gain reaches
$+11.2$--$+14.2$ points, largest on Entity ($+14.2$) and Long-Tail
($+12.8$). All improvements are significant (paired $t$-test,
$p<0.05$). INT4 quantization reduces recall by at most 0.87 points
($p>0.05$, not significant).

\subsubsection{Ranking Training Strategy Ablation}

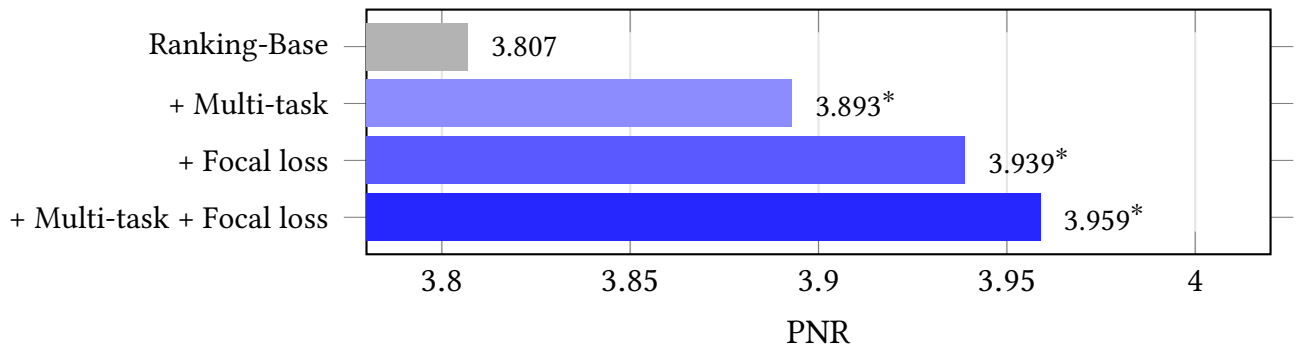
\begin{figure}[t]
\centering
\small
\begin{tikzpicture}
\begin{axis}[
    width=\dimexpr\columnwidth-26pt\relax,
    height=3.2cm,
    xbar, bar width=9pt,
    xmin=3.78, xmax=4.02,
    ymin=0.35, ymax=4.65,
    ytick={4,3,2,1},
    yticklabels={Ranking-Base, + Multi-task, + Focal loss,
                 + Multi-task + Focal loss},
    yticklabel style={font=\fontsize{6}{7}\selectfont},
    xlabel={PNR}, xlabel style={font=\fontsize{6.5}{8}\selectfont},
    xticklabel style={font=\fontsize{6}{7}\selectfont},
    xticklabel={\pgfmathprintnumber[fixed,precision=2]{\tick}},
    xmajorgrids=true, grid style={gray!20},
    nodes near coords,
    point meta=explicit symbolic,
    every node near coord/.append style={font=\fontsize{6}{7}\selectfont,
        anchor=west, xshift=1.5pt},
    clip=false,
]
    \addplot[fill=black!30, draw=none, bar shift=0pt]
        coordinates {(3.807,4) [3.807]};
    \addplot[fill=blue!45, draw=none, bar shift=0pt]
        coordinates {(3.893,3) [3.893$^{*}$]};
    \addplot[fill=blue!65, draw=none, bar shift=0pt]
        coordinates {(3.939,2) [3.939$^{*}$]};
    \addplot[fill=blue!85, draw=none, bar shift=0pt]
        coordinates {(3.959,1) [3.959$^{*}$]};
\end{axis}
\end{tikzpicture}
\caption{Training strategy ablation on the ranking benchmark.
Each bar adds the indicated strategy to Ranking-Base
(PNR annotated at the bar end).
$^\ast$: significant over Ranking-Base
(paired $t$-test, $p<0.05$).}
\label{fig:ranking_main}
\end{figure}

We start from Ranking-Base and evaluate each
training strategy both individually and in combination.
Figure~\ref{fig:ranking_main} reports PNR on the 400K human-annotated ranking
benchmark with satisfaction-oriented pairwise labels. \revfix{Offline we report only PNR, as answer-level evaluation is costly; PNR does not convert to the online GSB lift, but the two agree empirically, so PNR serves as its offline indicator.} The four bars correspond to
(i) Ranking-Base (PNR~$=$~3.807),
(ii) +Multi-task (PNR~$=$~3.893),
(iii) +Focal loss (PNR~$=$~3.939), and
(iv) +Multi-task + Focal loss (PNR~$=$~3.959).
Each bar independently adds the indicated strategy to Ranking-Base.
Both strategies combined achieve PNR~$=$~3.959
($+0.152$, $+4.0\%$, $p<0.05$), confirming complementary gains with
diminishing returns as the strategies overlap.

\subsection{Reinforced Selector Analysis}
\label{sec:experiments_selection}
The selector takes $\mathcal{D}_{\mathrm{rank}}$ and selects a subset $\mathcal{S}_q^*$ via dual-branch RL. We compare \revfix{Base, SFT, GRPO, and GRPO+OPD} on the same backbone and PE Judge. Table~\ref{tab:selection_ablation} reports \revfix{per-document rubric scores.}

\revfix{\textbf{Rubric-aligned training improves Utility from 9.45 for Base to 9.73 for GRPO+OPD ($+3.0\%$)}, while Authority, Freshness, and Quality remain stable. SFT supplies most of the gain (9.70); GRPO alone (9.68) stays below SFT, and only the combination (9.73) clears it. The gain is a rubric-conditioning effect: Base scores documents pointwise for relevance and authority and then applies submodular diversity re-ranking, so nothing in its objective rewards covering the query's rubric criteria, whereas teacher labels, SFT targets, and the set-level reward are defined over $\mathcal{R}_q$. Utility is near saturation, so the gain is modest but consistent. For training-dynamics details, see Appendix~\ref{sec:appendix_dynamics}.}

\begin{table}[t]
\centering
\caption{Selection ablation: per-document rubric scores. Utility shows the
clearest improvement ($+3.0\%$); Authority, Freshness, and Quality remain
stable.}
\label{tab:selection_ablation}
\small
\setlength{\tabcolsep}{3pt}
\begin{tabular}{l c c c c}
\toprule
\textbf{Method} & \textbf{Utility} & \textbf{Authority} & \textbf{Freshness} & \textbf{Quality} \\
\midrule
\revfix{Base}          & 9.45 & 9.64 & 9.81 & 9.47 \\
SFT                   & 9.70$^\ast$ & 9.63 & 9.86 & 9.55 \\
GRPO                  & 9.68$^\ast$ & 9.64 & 9.87 & 9.63 \\
\textbf{GRPO+OPD}     & 9.73$^\ast$ & 9.62 & 9.85 & 9.65 \\
\midrule
$\Delta$\% (vs.\ Base) & \textbf{+3.0\%}$^\ast$ & $-$0.2\% & +0.4\% & +1.9\% \\
\bottomrule
\end{tabular}

\vspace{2pt}
{\footnotesize
\revfix{Base is the production selector deployed on Yuanbao Search.}\\
$^\ast$: significant over Base (paired $t$-test, $p<0.05$).\par}
\end{table}

\section{Related Work}

\subsection{Neural Retrieval and Ranking}

\looseness=-1 Learning to rank has evolved from classical methods using hand-crafted
features~\cite{burges2010ranknet,yin2016ranking} to neural models that
learn semantic representations end to end.
Dense retrieval replaced sparse lexical matching with dual-encoder
architectures trained via contrastive learning with in-batch
negatives~\cite{karpukhin2020dpr,xiong2021ance}, and
ColBERT~\cite{khattab2020colbert,santhanam2022colbertv2} introduced late
interaction for token-level matching.
Encoder advances such as SimCSE~\cite{gao2021simcse},
Contriever~\cite{izacard2022contriever}, E5~\cite{wang2024e5},
BGE~\cite{chen2024bge}, and Qwen3-Embedding~\cite{zhang2025qwen3embed}
improved dense embeddings and scaled to multilingual settings, deployed with
Faiss~\cite{johnson2019faiss} and Matryoshka
representations~\cite{kusupati2022mrl}; ListT5~\cite{yoon2024listt5}
advanced listwise retrieval.
On the ranking side, LLM-based rankers have demonstrated strong zero-shot
and fine-tuned performance:
RankGPT~\cite{sun2023chatgpt}, RankLLaMA~\cite{ma2024rankllama},
PRP~\cite{qin2024prp}, and Setwise~\cite{zhuang2024setwise} pioneered LLM
listwise reranking, specialized open-weight rankers, pairwise prompting, and
setwise selection, respectively, and ReasonRank~\cite{liu2026reasonrank} adds
explicit reasoning; industrial web search has likewise benefited from
pre-trained language model rankers~\cite{zou2021plmranking}.
Focal loss~\cite{lin2017focal} has been adopted to address class imbalance
in ranking.
A common limitation across these methods is that they optimize for
per-document relevance, yet coverage, contradiction, and redundancy, which
emerge only when documents are combined into a set, lie outside the scope
of pointwise scoring.
We build on them by training retrieval and ranking with
rubric-grounded utility labels that encode set-level requirements (e.g.,
``covers both pros and cons'') into per-document supervision, aligning
early stages with the answer quality they serve.

\subsection{Document Set Selection and Agentic Search}

\looseness=-1 Beyond per-document scoring, explicit document subset composition has been
studied extensively.
Classical approaches reduce redundancy through submodular
optimization~\cite{lin2011submodular}, maximal marginal
relevance~\cite{carbonell1998mmr}, and determinantal point processes.
Information-gain-based methods, including InfoGain-RAG~\cite{wang2025infogainrag}
and Uplift-RAG~\revfix{\cite{qu2025upliftrag}}, select documents that maximize
knowledge gain for the generator.
Diversity-oriented approaches including CL4DIV~\cite{deng2024cl4div}
optimize for complementary coverage.
Generator-aware selectors directly optimize subsets for downstream answer
quality: SARA~\revfix{\cite{jin2025sara}} performs iterative evidence selection
through hybrid compression, R3~\cite{zhou2025r3} trains
reinforcement-learning-based retrievers, and
ReSearch~\cite{chen2025research} scales deep research through RL-driven
multi-turn search.
Agentic search further shifts retrieval from one-shot ranking to iterative
reasoning-and-retrieval loops:
ReAct~\cite{yao2023react} established the reasoning-and-acting paradigm;
Self-RAG~\cite{asai2024selfrag} introduced self-reflection to decide when
and what to retrieve; ALCE~\cite{gao2023alce} provides a systematic
framework for evaluating retrieval-augmented pipelines; and
LRAT~\cite{zhou2026lrat} trains retrievers from agent trajectories.
\looseness=-1 These methods focus on optimizing the selection stage, taking upstream
retrieval and ranking as given.
We instead train retrieval and ranking with the same rubric, so the
selector receives documents that already target set-level quality rather
than topical relevance.

\subsection{Rubric-Based Evaluation and Reinforcement Learning}

\looseness=-1 Rubrics decompose holistic quality judgments into discrete, verifiable
criteria.
LLM-based evaluation was established by
MT-Bench~\cite{zheng2023llmjudge} and advanced by
G-Eval~\cite{liu-etal-2023-g-eval} and
Prometheus~\cite{kim2024prometheus}, which improved alignment through
chain-of-thought scoring.
RUBRIC~\revfix{\cite{farzi2024rubric}} and LLM-Rubric~\cite{hashemi2024rubric}
applied criterion-decomposed evaluation to retrieval and
search-augmented generation.
Recent work has begun transforming rubrics from evaluation instruments
into training signals:
RaR~\revfix{\cite{gunjal2025rar}} uses checklist-style rubrics as GRPO
rewards, Chasing the Tail~\cite{zhang2026chasingtail} designs rubric reward
models for long-tail post-training, RuscaRL~\revfix{\cite{zhou2026ruscarl}}
breaks exploration bottlenecks via rubric-scaffolded RL, and Search
Wisely~\revfix{\cite{wu2025searchwisely}} reduces retrieval uncertainty.
On the RL front, PPO~\cite{schulman2017ppo} and
RLHF~\cite{ouyang2022instructgpt} established preference-based alignment,
DPO~\cite{rafailov2023dpo} simplified it through direct preference pairs,
GRPO~\cite{shao2025deepseekr1} enables efficient policy optimization via
within-group advantage estimation, and Scaf-GRPO~\cite{zhang2026scafgrpo}
adds scaffolded exploration.
GRPO-based selection methods typically rely on a single-branch design; our dual-branch extension (detailed in \S\ref{sec:selection}) adds best-rollout distillation for stable convergence under sparse set-level reward.
Prior work has established rubrics as evaluation protocols and applied
rubric rewards to individual LLM components. We instead use the rubric as a
shared supervision surface connecting retrieval, ranking, and selection under
one set-level utility objective, so each stage targets the answer quality it
produces.

\section{Conclusion and Future Work}

\looseness=-1 In this work, we present practical solutions to align the
search pipeline with rubric-grounded utility in a large-scale agentic
search engine.
The proposed solutions are successfully implemented and deployed to power
Yuanbao Search.
To bridge the granularity mismatch between set-level rubric evaluation
and per-document scoring, we proposed a lightweight labeler that distills
rubric-grounded judgments into per-document utility labels, supervising
a contrastive dual-encoder for utility-aligned retrieval and a
multi-task ranker with rubric-aligned utility scoring.
To manage the sparse reward and credit assignment problem in subset
optimization, we design a dual-branch RL selector that combines GRPO
with best-rollout distillation, using the best rollout as a
self-teacher for token-level correction.
Extensive offline and online experiments verify the
effectiveness of our proposed solutions.
Future work includes dynamically refreshing rubrics from fresh web
evidence to handle rapidly evolving topics, and extending the pipeline
to multimodal queries where criteria span text, images, and structured
data.

\clearpage
\begin{acks}
\revfix{We thank all members of the Yuanbao Search team, especially Jane~Xi
for resources and technical support.}
\end{acks}


\afterpage{%
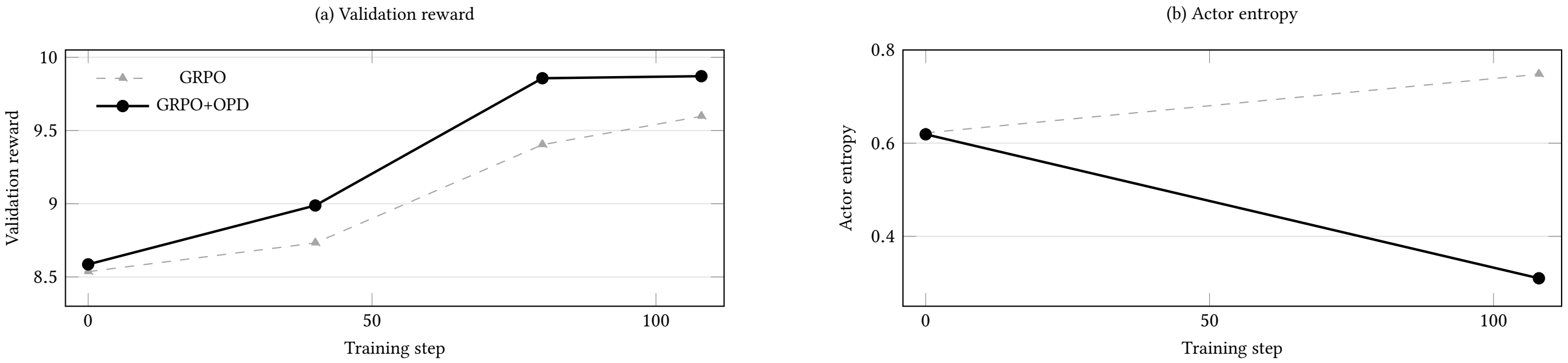
\begin{figure*}[t]
\centering
\begin{tikzpicture}
\begin{axis}[
  width=0.51\textwidth, height=4.5cm,
  xlabel={Training step}, ylabel={Validation reward},
  xmin=-4, xmax=112, ymin=8.3, ymax=10.05,
  xtick={0,50,100}, ytick={8.5,9,9.5,10},
  title={\scriptsize (a) Validation reward},
  tick label style={font=\scriptsize}, label style={font=\scriptsize},
  ymajorgrids=true, grid style={gray!20},
  legend style={font=\scriptsize, at={(0.03,0.97)}, anchor=north west,
                draw=none, fill=none},
]
\addplot[dashed, mark=triangle*, mark size=1.6pt, gray!70]
  coordinates {(0,8.536)(40,8.732)(80,9.403)(108,9.597)};
\addplot[thick, mark=*, mark size=1.6pt, black]
  coordinates {(0,8.586)(40,8.988)(80,9.857)(108,9.871)};
\legend{GRPO, GRPO+OPD}
\end{axis}
\end{tikzpicture}\hfill
\begin{tikzpicture}
\begin{axis}[
  width=0.51\textwidth, height=4.5cm,
  xlabel={Training step}, ylabel={Actor entropy},
  xmin=-4, xmax=112, ymin=0.25, ymax=0.8,
  xtick={0,50,100}, ytick={0.4,0.6,0.8},
  title={\scriptsize (b) Actor entropy},
  tick label style={font=\scriptsize}, label style={font=\scriptsize},
  ymajorgrids=true, grid style={gray!20},
]
\addplot[dashed, mark=triangle*, mark size=1.6pt, gray!70]
  coordinates {(0,0.622)(108,0.748)};
\addplot[thick, mark=*, mark size=1.6pt, black]
  coordinates {(0,0.619)(108,0.310)};
\end{axis}
\end{tikzpicture}
\caption{Selector training dynamics over 108 steps: (a) validation reward;
(b) actor entropy.}
\label{fig:training_dynamics}
\end{figure*}
}

\bibliographystyle{ACM-Reference-Format}
\bibliography{references}

\appendix

\section{Selector Training Dynamics}
\label{sec:appendix_dynamics}

We compare two selectors that share the same data, training recipe, and PE
Judge: GRPO, and GRPO+OPD (ours), which adds best-rollout distillation to GRPO.
Both are trained for 108 steps on 180K training queries, with G=8 rollouts per
query. We track two quantities. Validation reward is the PE Judge set-level
score (0--10) on held-out queries, measured at steps 0, 40, 80, and 108.
Actor entropy measures how concentrated the policy is, so lower values mean
sharper output; we report it at steps 0 and 108.
Figure~\ref{fig:training_dynamics} plots both quantities across the 108 training steps.

\newpage

Two observations follow. First, GRPO+OPD keeps improving while GRPO stalls.
Its validation reward rises from 8.586 at step 0 to 9.871 at step 108, against
8.536 to 9.597 for GRPO. Between steps 80 and 108, GRPO+OPD gains only 0.014
while GRPO still gains 0.194. Second, actor entropy separates the two
branches in the same direction: GRPO's entropy rises from 0.622 to 0.748 as the
policy drifts, while GRPO+OPD's falls to 0.310. GRPO+OPD also takes 608\,s
per step against 941\,s for GRPO. The final quality gap is small
(Table~\ref{tab:selection_ablation}, $+0.03$), so the branch's value is
faster, lower-entropy convergence rather than a large gain.

\balance

\end{document}